\documentclass[preprint,amsmath,amssymb,altaffillsymbol,nofootinbib,titlepage]{revtex4-1}
\usepackage{url}
\usepackage{color}
\usepackage{isotope}
\usepackage{pstricks}
\usepackage{colordvi}
\usepackage{epsfig}% Include figure files
\usepackage{graphicx}% Include figure files
\usepackage{dcolumn}% Align table columns on decimal point
\usepackage{bm}% bold math
\usepackage{subfigure}
\usepackage{amsmath}
\usepackage{amssymb}
\usepackage{hyperref}
\usepackage{slashed}
\usepackage{lineno}

\hypersetup{
    colorlinks=true, %set true if you want colored links
    linkcolor=blue,  %choose some color if you want links to stand out
   citecolor=blue
}

\def\gsim{\gtrsim}
\begin{document}

\title{Measurement of the Spectral Function of $\mathbf {^{40}}$Ar through \\the $\mathbf{(e,e^\prime p)}$ reaction \\
\vspace{0.5cm} \small{Proposal (PR12-14-012) submitted to the \\
Jefferson Lab Program Advisory Committee PAC 42 \\ July 2014 \vspace{0.5cm}}}

\author{O.~Benhar\footnote[1]{Spokesperson}, F.~Garibaldi and G.~M.~Urciuoli}
\affiliation{INFN and Department of Physics, ``Sapienza'' Universit\`a di Roma, I-00185 Roma, Italy}
\author{C.~Mariani\footnotemark[1]\footnote[2]{Contact Person, email: mariani@vt.edu}, C.-M.~Jen\footnotemark[1], J.~M.~Link, and M.~L.~Pitt}
\affiliation{Center for Neutrino Physics, Virginia Tech, Blacksburg, VA, 24061, USA}

\author{D.~B.~Day\footnotemark[1], D.~G.~Crabb, D.~Keller, R.~Lindgren, N.~Liyanage, O.~A.~Rondon, and J.~Zhang }\affiliation{ Department of Physics, University of Virginia, Charlottesville, VA, 22904, USA}
\author{D.~W.~Higinbotham\footnotemark[1], D.~Gaskell, O.~Hansen, C.~E.~Keppel, L.~Myers, B.~Sawatzky, and S.~Wood}\affiliation{Thomas Jefferson National Accelerator Facility, Newport News, VA, 23606, USA}
\author{A.~Ankowski and M.~Sakuda}\affiliation{Department of Physics, Okayama University, Okayama 700-8530, Japan}
\author{C. Giusti and A. Meucci}\affiliation{Department of Physics and INFN, University of Pavia,  I-27100 Pavia, Italy}
\author{G.~Garvey, X.~Jiang, and G.~B.~Mills}\affiliation{Los Alamos National Laboratory, Los Alamos, NM, 87545, USA}
\author{P.~A.~Souder, R.~Holmes, and R. Beminiwattha}\affiliation{Department of Physics, Syracuse University, Syracuse, NY, 13210, USA}
\date{\today}

\begin{abstract}
The interpretation of the signals detected by high precision experiments aimed at measuring neutrino oscillations requires an accurate description of the neutrino-nucleus cross sections. One of the key element of the analysis is the treatment of nuclear effects,  which is one of the main sources of systematics for  accelerator based experiments such as the Long Baseline Neutrino Experiment (LBNE). A considerable effort is currently being made to develop theoretical models capable of providing a fully quantitative description of the neutrino-nucleus cross sections in the kinematical regime relevant to LBNE. The approach based on nuclear many-body theory and the spectral function formalism has proved very successful in explaining the available electron scattering data in a variety of kinematical conditions. The first step towards its application to the analysis of neutrino data is the derivation of the spectral functions of nuclei employed in neutrino detectors, in particular argon. We propose a measurement of the coincidence $(e,e^\prime p)$ cross section on argon. This data will provide the experimental input indispensable to construct the argon spectral function, thus paving the way for a reliable estimate of the neutrino cross sections. In addition, the analysis of the $(e,e^\prime p)$ data will help a number of theoretical developments, like the description of final-state interactions needed to isolate the initial-state contributions to the observed single-particle peaks, that is also needed for the interpretation of the signal detected in neutrino experiments.
\par We request 9 days of beam time at 2.2~GeV, 1 day for calibration and 8 days for the measurement of parallel and anti-parallel kinematics of $(e,e^\prime p)$ on an argon target. The beam time request accounts for radiative losses and includes several hours of data taking dedicated to measure possible backgrounds. This measurement will provide the only available 
high statistics sample of electron scattering data on argon in reduced-FSI kinematics.
\end{abstract}

\maketitle

\tableofcontents

\newpage

\section{M\lowercase{otivation}}\label{intro}

Neutrino physics is entering the age of precision measurements. Several experiments have detected neutrino oscillations, thus providing unambiguous evidence  that neutrinos---assumed to be massless in the Standard Model of particle physics---have in fact non-vanishing masses.
Reactor neutrino experiments carried out in the last five years (Double Chooz \cite{DC}, Daya Bay \cite{DB} and RENO \cite{RENO}) recently reported high quality measurements of the $\theta_{13}$ mixing angle,
%that turned out to be quite large, its value being $\sim 10 \ {\rm deg}$.
the value of which turned out to be $\sim$10 deg.
The large $\theta_{13}$ mixing angle will enable future experiments -- such as the Long-Baseline Neutrino Experiment (LBNE) in the United
States~\cite{LBNE}---to search for leptonic CP violation in appearance mode, thus addressing one of the outstanding problems in particle physics. However, these searches will involve high precision determinations of the oscillation parameters, which in turn require a deep understanding  of neutrino interactions with matter.
In view of the achieved and expected experimental accuracies, the treatment of nuclear effects is indeed regarded as one of the main sources of systematic uncertainty.

Over the past decade, it has become more and more evident that the independent particle model of nuclei---the ultimate implementation of which is the relativistic Fermi gas model (RFGM) routinely employed in simulation codes---is not adequate to account for the complexity of nuclear dynamics and the variety of reaction mechanisms contributing to the detected signals.

The large discrepancy between the results of Monte Carlo simulations and the double differential cross section of charged current (CC) quasielastic (QE) interactions in carbon, recently measured by the MiniBooNE Collaboration using a beam of average energy $\sim$0.8 (0.7) GeV in the neutrino (antineutrino) mode, is a striking manifestation of the above problem~\cite{MBCCQE,MBNC,ref:MB_anu}. More recently, the analysis of the inclusive $\nu_\mu$-nucleus cross sections at beam energy in the range $2-20$ GeV, measured by the 
MINER$\nu$A Collaboration using a variety of targets, led to the conclusion that none of nuclear models implemented in Monte Carlo simulations is capable of reproducing the data ~\cite{ref:MINERvA_ratios}.

A considerable effort is currently being made to develop theoretical models capable of providing a fully quantitative description of the neutrino-nucleus cross section in the kinematical regime relevant to LBNE, corresponding to beam energies ranging from a few hundred MeV to a few GeV. In this context, a key element is the information provided by the large body of theoretical and experimental studies of electron-nucleus scattering.

The approach based on nuclear many-body theory and the spectral function formalism has proved very successful in explaining the available electron scattering data in a variety of kinematical conditions (for a recent review on the quasielastic sector, see Ref.~\cite{RMP}). The first step in its application to the analysis of neutrino data is the derivation of the spectral functions of nuclei employed in neutrino detectors.

The spectral function, $P({\bf k},E)$,  yields the probability of removing a nucleon of momentum~${\bf k}$ from the nuclear ground state leaving the residual system with excitation energy $E$. Accurate theoretical calculations of $P({\bf k},E)$, based on nuclear many-body theory, have
been carried out for the three-nucleon system~\cite{ingo,sauer} and uniform, isospin-symmetric, nuclear matter~\cite{pke_nm}. For the isospin symmetric $p$-shell  nuclei $^{16}$O and $^{12}$C, relevant to water Cherenkov detectors (e.g. Super-K) and mineral-oil detectors (e.g. MiniBooNE), respectively, the spectral functions  have been obtained in Refs.~\cite{LDA,PRD}, combining theoretical calculations and the information provided
by coincidence $(e,e^\prime p)$ experiments. Within this scheme, the availability of the $(e,e^\prime p)$ data is {\em essential} to accurately describe
binding energies, spectroscopic factors, and widths of the shell model states.

As future neutrino experiments---most notably LBNE---will use large liquid argon detectors to perform a precision measurement of the CP violating phase, understanding the response of argon to neutrino and antineutrino interactions is of paramount importance. To obtain the spectral functions of calcium $^{40}$Ca and argon $^{40}$Ar, the authors of Ref.~\cite{ANK} have proposed a model, inspired by the one developed in Refs.~\cite{LDA,PRD} but involving additional and rather crude approximations.

The electron scattering data for calcium is scarce. A $(e,e^\prime p)$ experiment with low missing energy resolution  has been carried out at Saclay in the 1970s~\cite{40CaSaclay}, while the more recent, high resolution, measurements performed at NIKHEF-K~\cite{40CaNIKHEF} cover a limited energy range. The calcium spectral function of Ref.~\cite{ANK} has been derived using the results of theoretical calculations of the momentum distribution~\cite{bisconti} and the empirical energy spectrum predicted by the dispersive optical-model analysis of Ref.~\cite{mahaux}.

As far as argon is concerned, the current state of affairs looks even more problematic, the only available electron scattering data being the {\em inclusive} cross sections measured at Frascati in the 1990s~\cite{LNF2}.  The empirical information on the energy levels is limited to a~few neutron states in the vicinity of the Fermi surface, obtained in Ref.~\cite{Arspec}, and no theoretical calculations of the momentum distribution are available. Owing to the lack of the needed input information  the derivation of a model spectral function, even within the oversimplified scheme proposed in Ref.~\cite{ANK}, involves very large uncertainties.

The inclusive electron-argon cross section obtained using the spectral function of Ref.~\cite{ANK} turns out to be in fairly good agreement with the Frascati data.
The same spectral function has been also used to obtain $\nu_\mu$ and $\bar\nu_\mu$ CC cross sections consistent with the measurements  recently  reported by the ArgoNeuT Collaboration, obtained at  mean neutrino (antineutrino) energy of 9.6 (3.6) GeV~\cite{ref:ArgoNeuT}.

However, a more realistic model---including a more accurately determined spectrum and measured spectroscopic factors and momentum distributions of the shell model states---will be needed to reach an accuracy comparable to that required for precise neutrino experiments, as well as to describe more
exclusive processes.

Owing to the strong isospin dependence of nuclear forces, nucleon-nucleon interactions in the proton-neutron sector are very different from  those  taking place in the proton-proton and neutron-neutron channels. As a consequence, in the absence of both experimental information on argon spectroscopic factors and theoretical input on the correlation contribution to the spectral function of isospin-asymmetric nuclear matter, $P({\bf k},E)$ cannot be accurately determined within the scheme of Refs.~\cite{LDA, PRD}.

A direct measurement of the coincidence $(e,e^\prime p)$ cross section on argon would provide the experimental input indispensable to constrain
the generalization of the theoretical model presented in Ref.~\cite{LDA, PRD} to argon, thus paving the way for a reliable estimate of the neutrino cross sections. In addition, the analysis of the $(e,e^\prime p)$ data will require a number of theoretical developments---e.g. the description of final-state interactions needed to isolate the initial-state contributions to the observed single-particle peaks~\cite{FSI}---that will also be needed for the interpretation of the signal detected in neutrino experiments.

To see how the description of nuclear dynamics affects the measurement of neutrino oscillations, consider the simplest 
case of two neutrino flavors, $\alpha$ and $\beta$. The oscillation probability 

\begin{equation}
P_{\alpha \rightarrow \beta} = \sin^2(2\theta)\sin^2\left(\frac{\Delta m^2 L}{4E_\nu}\right) \ , 
\label{eq:oscillation_probability}
\end{equation}
~~\\
\noindent
driven by a single mixing angle $\theta$ and the squared-mass difference $\Delta m^2$, is a function of the neutrino energy $E_\nu$. 
Accelerator experiments obtain neutrino beams as decay products mainly of pions, which are in turn produced by interactions of the primary proton beam with the 
target material. As a consequence, the neutrino beam is not monochromatic, and in any interaction events the neutrino energy must be inferred from  
the measured kinematical variables of the particles involved the process.

In the case of CC QE scattering off a free nucleons at rest, the observed energy and production angle of the outgoing charged lepton provides sufficient information to determine $E_\nu$. However, this is no longer the case for CC QE scattering off nuclear targets, the use of which is mandatory to reach acceptable statistics. While typically the neutrino energy is still reconstructed from the measured kinematics of the charged lepton only, the accuracy of this method is limited by the accuracy to which nuclear effects are described by the Monte Carlo simulations employed for data analysis.

The distributions of nucleon's momentum and energy--described by the nuclear spectral function--smear the reconstructed energy. In addition, the interaction products undergo 
final state interactions with the surrounding nucleons, which also affects their energy. Finally, in the nuclear medium more complex reaction mechanisms, such as those involving two-body currents, have to be taken into account, as they significantly affect energy reconstruction.

The authors of Ref.~\cite{ref:Mosel_ERec} have shown that the irreducible background arising from pion production followed by nuclear absorption also plays a critical role, 
and may lead to underestimating  the neutrino energy by as much as $\sim$300~MeV. As a consequence, an accurate estimate and subtraction 
of this background from the CC QE event candidates is called for.

The above issues are all very relevant, as many experiments exploit QE interactions for their neutrino oscillation analysis. The efficiency of identification of QE interactions is usually very high, but there are other neutrino interactions (1$\pi$ and DIS) occurring at higher neutrino energies, with much lower reconstruction efficiency, due to the high multiplicity of 
the final state. 
Most of the current experiments rely on Monte Carlo predictions rather than direct observation to identify and quantify the efficiency and energy reconstruction bias arising from such 
non-QE events. The bias induced in the reconstructed neutrino energy by non-QE events may in fact be the source of the low energy excess observed by MiniBooNE.

The smearing, or the shift, of the reconstructed energy induced by non-QE interactions has been recently analyzed in 
Refs.~\cite{Coloma:2012ji,Lalakulich:2012hs,Martini:2012fa,Coloma:2013rqa,Coloma:2013tba}. Of particular interest are the studies of  Ref.~\cite{Coloma:2013rqa,Coloma:2013tba},  in which an ideal neutrino oscillation experiment is simulated using different nuclear models in the event generation and oscillation fit. 
Based on these analyses, the authors  conclude that nuclear models matter at the level of 10--20\% for the determination of the oscillation parameters. 
Of course, it should be considered that all recent neutrino experiments, such as MiniBooNE, MINOS, T2K and NO$\nu$A, have their systematic uncertainties on neutrino cross sections and nuclear models greatly reduced by a variety of cross-section measurements carried out using carbon and oxygen. However, the situation will be different, and much worse, in the case of LBNE,  which will use Argon as the target nucleus. 

As a final remark, we note that,  in addition to the importance for the neutrino program, the precise determination of the structure of the argon nucleus will be  
relevant to the LBNE search for nucleon decay leading to a violation of baryon-number conservation. For liquid-argon detectors, the channels of particular interest 
are the $p\rightarrow K^++\bar\nu$ and $p\rightarrow K^0+\mu+$. For a decaying proton at rest, the produced kaon would be monochromatic. However, the initial distribution of proton's momentum and energy in the nucleus--dictated by the spectral function--is responsible a non-negligible smearing of the kaon's momentum~\cite{ref:Argon_pDecay,ref:Omar_pDecay}. 
To achieve an accurate modeling of this nuclear effect, it is essential to obtain the relevant nuclear structure information measuring the $(e,e^\prime p)$ cross section off argon.

\section{The $(e,e^\prime p)$ cross section}\label{xsec}

We plan to study the process
\begin{align}
\label{process}
e + A \to e^\prime + (A-1)^\star + p \ ,
\end{align}
in which an electron of initial four-momentum $k_e\equiv(E_e,{\bf k}_e)$ scatters off a nuclear target to a state of four-momentum $k^\prime_e\equiv(E_{e^\prime},{\bf k}_{e^\prime})$. The hadronic final state consists of a proton of four momentum $p\equiv(E_p,{\bf p})$ and the undetected $(A-1)$-nucleon recoiling system.

The differential cross section of process \eqref{process} can be written in the from
\begin{align}
\frac{d\sigma_A}{dE_{e^\prime} d\Omega_{e^\prime} dE_p d\Omega_{p}} \ = {\rm K} \ \frac{\alpha^2}{Q^4} \ \frac{E_{e^\prime}}{E_e}
L^{\lambda \mu} W_{\lambda \mu}\ ,
\end{align}
where $K =  |{\bf p}| E_p$, while $\Omega_{e^\prime}$ and  $\Omega_{p}$ denote the solid angles specifying the directions of the outgoing electron and proton, respectively.
In the above equation $\alpha = 1/137$ is the fine structure constant, while the squared four momentum transfer is given by
\begin{align}
q^2 =  -Q^2 = \omega^2 - |{\bf q}|^2  \ ,
\end{align}
with $q = k_e - k^\prime_e \equiv(\omega,{\bf q})$.

The tensor $L^{\lambda \mu}$, that can be written, neglecting the electron mass, as
\begin{align}
L^{\lambda \mu} = 2 \left[ k_e^\lambda k_{e^\prime}^\mu + k_e^\mu k_{e^\prime}^\lambda
 - g^{\lambda\mu} (k_e k_{e^\prime}) \right] \ ,
\label{lepten}
\end{align}
with $g^{\lambda \mu} =  {\rm diag}(1,-1,-1,-1)$, is fully specified by the measured electron kinematical variables. All the information on the internal structure of the target is contained in the response tensor
\begin{align}
\label{response}
W_{\lambda \mu}= \langle 0 |J_\lambda| f  \rangle\langle  f |J_\mu|0\rangle
\delta^{(4)}(p_0+q-p_f)\ ,
\end{align}
the definition of which involves the initial and final nuclear states $|0 \rangle$ and $| f \rangle$, carrying four-momenta $p_0$ and $p_f$,
as well as the nuclear current operator,
\begin{align}
\label{def:current}
J^\mu = \sum_{i=1}^A j^\mu_{i} + \sum_{j>i=1}^A j^\mu_{ij} \ .
\end{align}
In the above equation, $j^\mu_i$ is the current describing the electromagnetic interaction of a single nucleon, while the operator $j^\mu_{ij}$ takes into account processes involving two nucleons, such as those associated with meson-exchange currents (MEC) \cite{MEC}.
\subsection{Plane Wave Impulse Approximation}\label{PWIA}

\begin{figure}[ht!]
\includegraphics[scale=1.2]{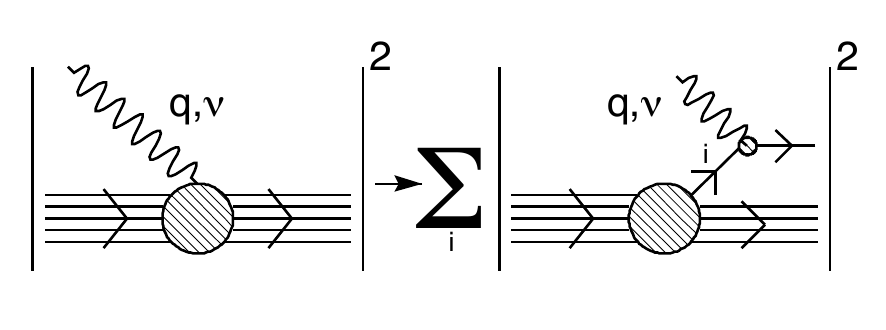}
\vspace*{-.3in}
\caption{Schematic representation of the IA regime, in which the nuclear cross section is replaced by the incoherent sum of cross sections describing scattering off individual nucleons, the recoiling $({\rm A}-1)$-nucleon system acting as a spectator.}
\label{IA:cartoon}
\end{figure}

The  Plane Wave Impulse Approximation (PWIA) is based on the assumptions that i) as the space resolution of the electron beam is $\sim 1/|{\bf q}|$, at large momentum transfer scattering off a nuclear target reduces to the incoherent sum of elementary scattering processes involving individual {\it bound} nucleons (see Fig.~\ref{IA:cartoon}) and ii) final state interactions between the hadrons produced at the  electron-nucleon vertex and the recoiling nucleus are negligibly small.

%%%%%%%%%%%%%%%%%%%%%%%%%%%%%%%%%%%%%%%%%%%%%%%%%%%%%%%%%%%%%%%%%%%%%%%%%%%%
%%%%%%%%%%%%%%%%%%%%%%%%%%%%%%%%%%%%%%%%%%%%%%%%%%%%%%%%%%%%%%%%%%%%%%%%%%%%

The above assumptions are implemented disregarding the contribution of the two-body current in Eq.~\eqref{def:current}, i.e. setting
\begin{align}
J^\mu \approx \sum_{i=1}^A j^\mu_i \ ,
\end{align}
and writing the nuclear final state appearing in Eq.~\eqref{response} in the factorized form
\begin{align}
\label{def:finalstate}
| f  \rangle = {\rm e}^{i {\bf p} \cdot {\bf r_1}} \ \eta(1) \ \Phi_f(2, \ldots ,A)  \ ,
\end{align}
where $\eta(1)$ describes the spin of the knocked out nucleon, labeled by the index $1$, while $\Phi_f(2, \ldots ,A)$ is the wave function associated with the recoiling nucleus.

Within the PWIA scheme, the $(e,e^\prime p)$ cross section reduces to the simple form
\begin{align}
\label{dsigma_PWIA}
\frac{d\sigma_A}{dE_{e^\prime} d\Omega_{e^\prime} dE_p d\Omega_{p} }  = {\rm K} \sigma_{ep} P(p_m, E_m )\ ,
\end{align}
with $p_m = |{\bf p}_m|$. The missing momentum, ${\bf p}_m$, which in the PWIA scheme can be identified with the initial momentum of the struck nucleon, is given by
\begin{align}
\label{def:pmiss}
{\bf p}_m = {\bf p}-{\bf q} \ ,
\end{align}
whereas the missing energy $E_m$, yielding its removal energy, is defined through
\begin{align}
\label{def1:Emiss}
\omega + M_A = \sqrt{ (M_A - m + E_m)^2 + |{\bf p}_m|^2 } + E_p \ ,
\end{align}
$m$ and $M_A$ being the proton and target masses, respectively.
Note that, in the limit of low missing momentum, $ |{\bf p}_m|^2/(M_A-m + E_m)^2 \to 0$, the expression of the missing energy reduces to the simple form
\begin{align}
\label{def2:Emiss}
E_m = \omega - T_p \ ,
\end{align}
where $T_p = E_p - m$ is the kinetic energy of the outgoing nucleon.

The probability distribution of finding a nucleon with momentum $p_m$ and removal energy $E_m$ in the target ground state is described by the spectral function, defined as
\begin{align}
\label{def:PkE}
P(p_m,E_m) = \sum_f  \left| \int \ d^3 r_1 \  {\rm e}^{i {\bf p}_m \cdot {\bf r}_1} \    \Psi({\bf r}_1 \ldots {\bf r}_A)^\dagger \Phi_f({\bf r}_2 \ldots {\bf r}_A) \right|^2  \ \delta(E_m + E_0 - E_f) \ ,
\end{align}
where $E_0$ is the target ground state energy and the sum includes all states of the recoiling nucleus, the energy of which is denoted $E_f$.

The kinematical region corresponding to low missing momentum and energy, where shell model dynamics dominates, has been extensively
studied by coincidence $(e,e^\prime p)$ experiments (for a review, see, e.g., Ref.\cite{book}). The spectral function extracted from the data
is usually written in the form
\begin{align}
P_{MF}(p_m,E_m) = \sum_{\alpha} Z_\alpha\ |\xi_\alpha(p_m)|^2 F_\alpha(E_m-E_\alpha) \ ,
\label{S:MF}
\end{align}
where the sum is extended to all occupied states belonging to the Fermi sea. Comparison to Eq. \eqref{def:PkE} shows that
 $\sqrt{Z_\alpha}  \ \xi_\alpha(p_m)$ is the Fourier transform of the overlap function
\begin{align}
\label{overlap}
\xi_\alpha({\bf r}_1) = \int d^3 x \ \Psi({\bf r}_1 \ldots {\bf r}_A)^\dagger \Phi_\alpha({\bf r}_2 \ldots {\bf r}_A) \ ,
\end{align}
where $\Phi_\alpha$ denotes the wave function  of the $(A-1)$-nucleon system with a hole in the state $\alpha$. The spectroscopic factors $Z_\alpha < 1$ and the functions $F_\alpha(E_m-E_\alpha)$, describing the energy width of the state of the outgoing nucleon, account for the effects of residual interactions not included in the mean field picture. In the absence of these interactions, the overlap of Eq.~\eqref{overlap} can be identified with the shell model wave function of the state $\alpha$,  $\phi_\alpha$, while $Z_\alpha \rightarrow 1$, and $F_\alpha(E_m-E_\alpha) \rightarrow \delta(E_m-E_\alpha)$.

In addition to the contribution of Eq.~\eqref{S:MF}, corresponding to processes in which the residual nucleus is left in a bound state, the spectral function includes contributions associated with final states in which one, or more, spectator nucleons are excited to the continuum. These contributions, the occurrence of which is a clear manifestation of nucleon-nucleon correlations in the initial state, have been computed for uniform nuclear matter, in a broad range of densities, using Correlated Basis Function (CBF) perturbation theory~\cite{LDA}. Within the Local Density Approximation (LDA), the nuclear matter results can be used to obtain the correlation contribution to the spectral function of a finite nucleus of mass number $A$ from
\begin{align}
P_{\rm corr}(p_m,E_m) = \int d^3r\ \rho_A({\bf r})
P^{NM}_{\rm corr}(p_m,E_m;\rho = \rho_A({\bf r})) \ ,
\label{S:corr}
\end{align}
where $\rho_A({\bf r})$ is the nuclear density distribution and $P^{NM}_{\rm corr}(p_mE_m;\rho)$ is the correlation part of the spectral function of uniform nuclear matter at density $\rho$.

The full LDA spectral function is given by the sum
\begin{align}
P(p_m,E_m) = P_{MF}(p_m,E_m) + P_{\rm corr}(p_m,E_m) \ .
\end{align}

The tenet underlying the LDA approach is that the correlation structure of the nuclear ground state, being mainly driven by short-range dynamics, is largely unaffected by surface and shell effects. This assumption is strongly supported by the predictions of a number of  theoretical calculations of the nucleon momentum distribution
\begin{align}
\label{def:momdist}
n(p_m) = \int dE \ P(p_m,E_m) \ ,
\end{align}
showing that for $A\ge$4 the quantity $n({\bf k})/A$ becomes nearly independent of $A$ at large momentum, typically $p_m \gsim 300$ MeV (see, e.g., Ref.~\cite{RMP2}).

%%%%%%%%%%%%%%%%%%%%%%%%%%%%%%%%%%%%%%%%%%%%%%%%%%%%%%%%%%%%%%%%%%%%%%%%%%%%		
\subsection{Final State Interactions}\label{FSI}

While the PWIA description provides a clear understanding of the mechanism driving the $(e,e^\prime p)$ reaction, corrections arising from Final State Interactions (FSI) between the outgoing nucleon and the residual nucleus are known to be in general not negligible, and must be carefully taken into account.

Within the Distorted Wave Impulse Approximation (DWIA), widely and successfully employed to analyze the large data set of $(e,e^\prime p)$
cross sections, the plane wave describing the motion of the outgoing nucleon is replaced by a scattering wave function $\chi_p$ which, in principle,  is an eigenfunction of the nonlocal Feshbach hamiltonian ${\mathcal H}$~\cite{FSI}. In the presence of FSI, the initial momentum of the hit nucleon, ${\bf p}_i$, cannot be trivially reconstructed through identification with the measured missing momentum.

The problem of obtaining $\chi_p$ can be greatly simplified by approximating ${\mathcal H}$ with a phenomenological optical potential, describing the interactions with the mean field of the residual nucleus. The optical potentials, determined fitting the available proton-nucleus scattering data, typically include complex central and spin-orbit components, as well as a Coulomb term.

Within the DWIA scheme, the nuclear cross section can no longer be written in the simple form of Eq.~\eqref{dsigma_PWIA}.
However, neglecting the effect of the spin-orbit potential, one can still recover a factorized expression in terms of the distorted spectral function
[compare to Eq.\eqref{S:MF}]
\begin{align}
P^D_{MF}({\bf p}_m,{\bf p},E_m) = \sum_\alpha Z_\alpha  | \xi^D_\alpha({\bf p}_m,{\bf p})|^2 F_\alpha(E_m-E_\alpha)  \ ,
\end{align}
with
\begin{align}
\sqrt{Z_\alpha} \ \xi^D_\alpha({\bf p}_m,{\bf p}) = \int d^3 p_i \ \chi_p^\star({\bf p}_i + {\bf q})  \xi({\bf p}_i) \ ,
\end{align}
The accuracy of the factorization scheme, which proves to be most useful for the extraction of the spectral function 
from the data, can be quantitatively tested comparing the resulting cross section with that obtained from a direct
many-body calculation of the relevant nuclear transition matrix elements. 

The large body of existing work on $(e,e^\prime p)$ data suggests that the effects of FSI can be strongly reduced measuring the cross
section in kinematical conditions such that ${\bf p} \parallel {\bf q}$. For $|{\bf p}| > |{\bf q}$ ($|{\bf p}| < |{\bf q}$) , implying in turn
$p_m >0$ ($p_m <0$),  this setup is referred to as parallel (antiparallel) kinematics \cite{DWIA1}.

In parallel kinematics, the distorted momentum distribution at fixed $|{\bf p}|$, corresponding to fixed energy of the detected proton, becomes a
function of missing momentum only
\begin{align}
N^D_\alpha(p_m) = Z_\alpha \ | \xi^D(p_m) |^2 \ ,
\end{align}
and the effects FSI can be easily identified. The real part of the optical potential  brings about a shift in missing momentum, while inclusion of the
the imaginary part leads to a significant  reduction of the PWIA result, typically by a factor $\sim 0.7$.

The treatment of FSI outlined in the is Section, developed by the Pavia Group, has been extensively tested, and employed to extract nuclear spectral functions  from the 
$(e,e^\prime p)$ cross sections measured at Saclay and NIKHEF. It has been recently applied to neutrino-argon interactions, in a study of the inclusive 
cross section measured by the ArgoNeuT Collaboration\cite{pavia_argoneut}.

In addition to hadronic FSI, in a nucleus as heavy as argon the distortion of the electron wave functions arising from interactions with the Coulomb field 
of the nucleus are also expected to be non negligible.  Their effect, that can be taken into account through an expansion in powers of $Z\alpha$, has been analyzed in 
Ref.~\cite{distortion}. The results of this study indicate that in the case of $^{40}{\rm Ca}$ the simple prescription based on the use of an effective momentum transfer
provides remarkably accurate results. 

%%%%%%%%%%%%%%%%%%%%%%%%%%%%%%%%%%%%%%%%%%%%%%%%%%%%%%%%%%%%%%%%%%%%%%%%%%

\section{Relevant kinematical domain}\label{kinematics}

We plan to study the $(e,e^\prime p)$ cross sections in the kinematical region in which single nucleon knock out of a nucleon
occupying a shell model orbit is the dominant reaction mechanism. The relevant missing energy domain is illustrated in Tab.~\ref{energy_spectra}, listing the separation energies of the proton and neutron shell model states belonging to the Fermi sea of $\isotope[40][18]{Ar}$~\cite{ANK}.
For comparison, the corresponding quantities for $\isotope[40][20]{Ca}$ ground state are reported in the same table.

%%%%%%%%%%%%%%%%%%%%%%%%%%%%%%%%%%%%%%%%%%%%%%%%%%%%%%%%%%%%%%%%%%%%%%%%%%%%%%%%%
\begin{table}[ht!]
\begin{center}
\begin{tabular}{ c c c c c   }
\hline
\hline
 & \multicolumn{2}{d}{\mspace{25mu}\text{protons}} & \multicolumn{2}{d}{\mspace{25mu}\text{neutrons}}\\
% \hline
 & \multicolumn{1}{c}{$\isotope[40][20]{Ca}$} & {$\isotope[40][18]{Ar}$} & \multicolumn{1}{c}{$\isotope[40][20]{Ca}$} & {$\isotope[40][18]{Ar}$}\\

% $\rho \  [{\rm fm}^{-3}]$                    & $F_0$ &  $F_1$  & $F_2$ &  $G_0$  &  $G_1$ & $G_2$\\
% $[{\rm fm}^{-3}]$  &          &             &           &              &            & \\ \\
\hline \hline
$1s_{1/2} \ \ \ $ & 57.38   & 52  &   66.12   & 62 \\
$1p_{3/2} \ \ \ $ & 36.52  & 32   &  43.80    & 40  \\
$1p_{1/2} \ \ \ $ & 31.62  & 28   &  39.12    & 35  \\
$1d_{5/2} \ \ \ $ & 14.95  & 11   &  22.48    & 18  \\
$2s_{1/2} \ \ \ $ & 10.67  &   8   &  17.53    & 13.15 \\
$1d_{3/2} \ \ \ $ & 8.88   &    6   &  15.79   & 11.45  \\
$1f_{7/2} \ \ \ $ &            &         &             & 5.56 \\
\hline
\hline
\end{tabular}
\caption{\label{energy_spectra} Separation energy of the proton and neutron shell model orbits relevant to the $\isotope[40][20]{Ca}$ and $\isotope[40][18]{Ar}$ ground states (adapted from Ref.~\cite{ANK}).}
\end{center}
\end{table}
%%%%%%%%%%%%%%%%%%%%%%%%%%%%%%%%%%%%%%%%%%%%%%%%%%%%%%%%%%%%%%%%%%%%%%%%%%%%%%%%%

%%%%%%%%%%%%%%%%%%%%%%%%%%%%%%%%%%%%%%%%%%%%%%%
\begin{figure}[ht!]
%\vspace*{.1in}
 \includegraphics[scale= 0.60]{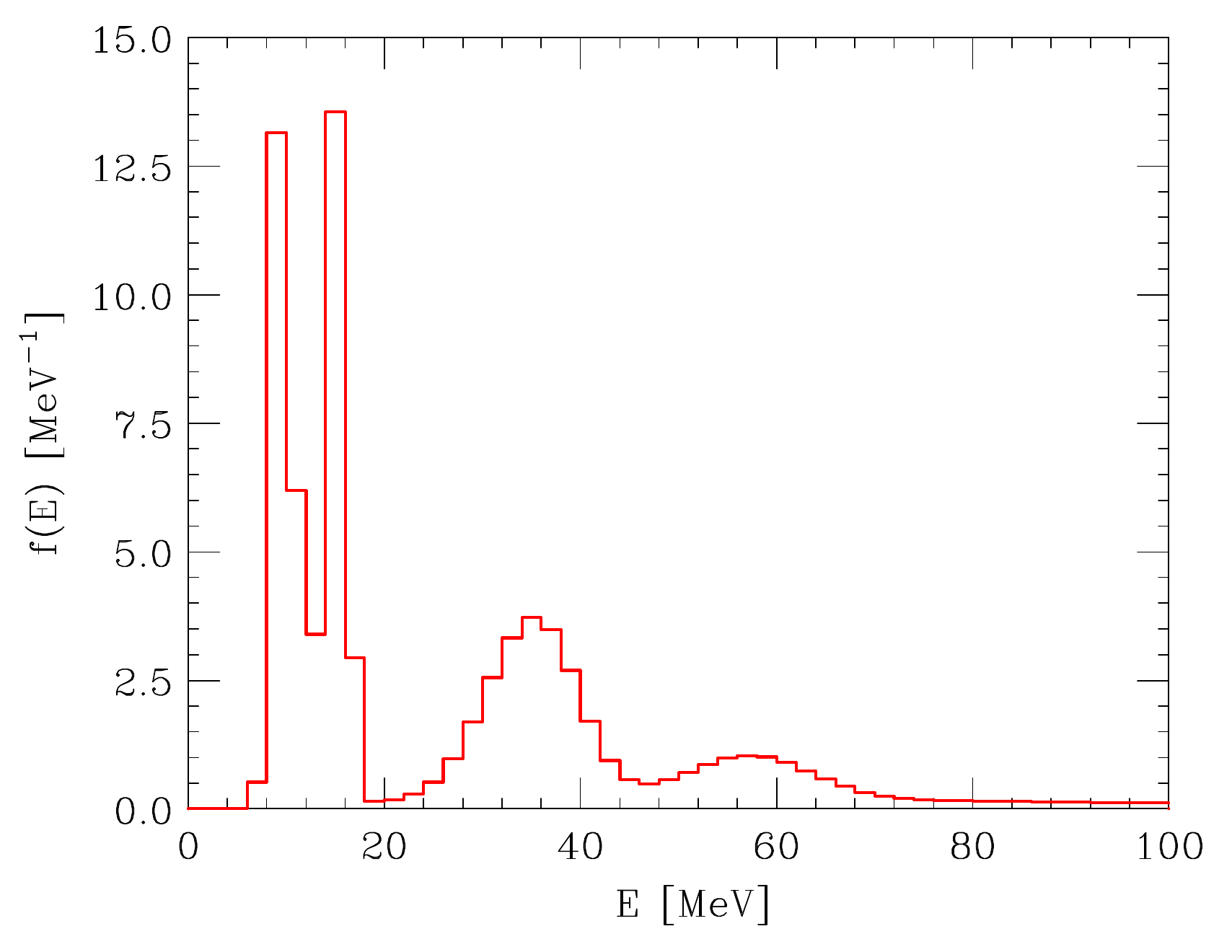}
\caption{Proton removal energy distribution in the $\isotope[40][20]{Ca}$ ground state, obtained from Eq.\eqref{edist} using the spectral function model of Ref.~\cite{ANK}. 
Note that the experimental missing energy resolution is expected to be better than that shown in this Figure.  }
\label{edist_ca}
\end{figure}
%%%%%%%%%%%%%%%%%%%%%%%%%%%%%%%%%%%%%%%%%%%%%%%%%%%%%%%%%%

Figure~\ref{edist_ca} shows the energy distribution
\begin{align}
\label{edist}
{\rm f}(E) = 4 \pi \int dk \ k^2 P(k,E) \,
\end{align}
obtained from the model proton spectral function of $\isotope[40][20]{Ca}$ discussed in Ref.~\cite{ANK}. The spectroscopic lines corresponding to the valence orbitals are clearly visible, as are the bumps associated with more deeply bound states. Note, however, that the experimental missing energy resolution is expected to be better that that 
illustrated in Fig.~\ref{edist_ca}.  

{\bf The results of Tab.~\ref{energy_spectra} and Fig.~\ref{edist_ca} suggest that the investigation of the mean field contribution to the Argon spectral function will require a scan of the missing energy domain extending from $E_m \sim$ 8~MeV to $E_m \sim$ 60~MeV, a $E_m$ range that can easily be achieved in (e,e$^\prime$p) experiments with the Hall A HRS spectrometers.}

\begin{figure}[ht!]
%\vspace*{.1in}
 \includegraphics[scale= 0.46]{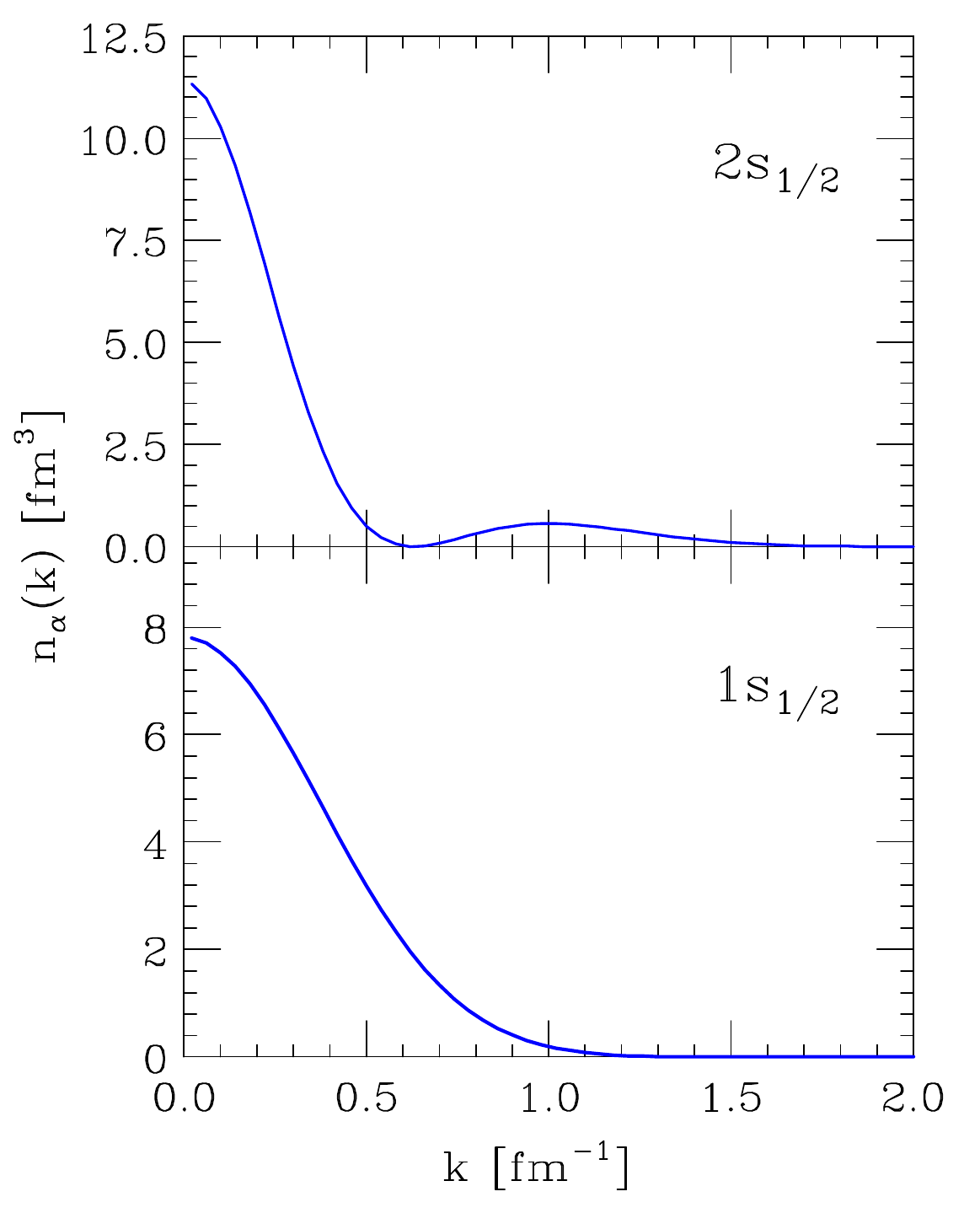}  \includegraphics[scale= 0.46]{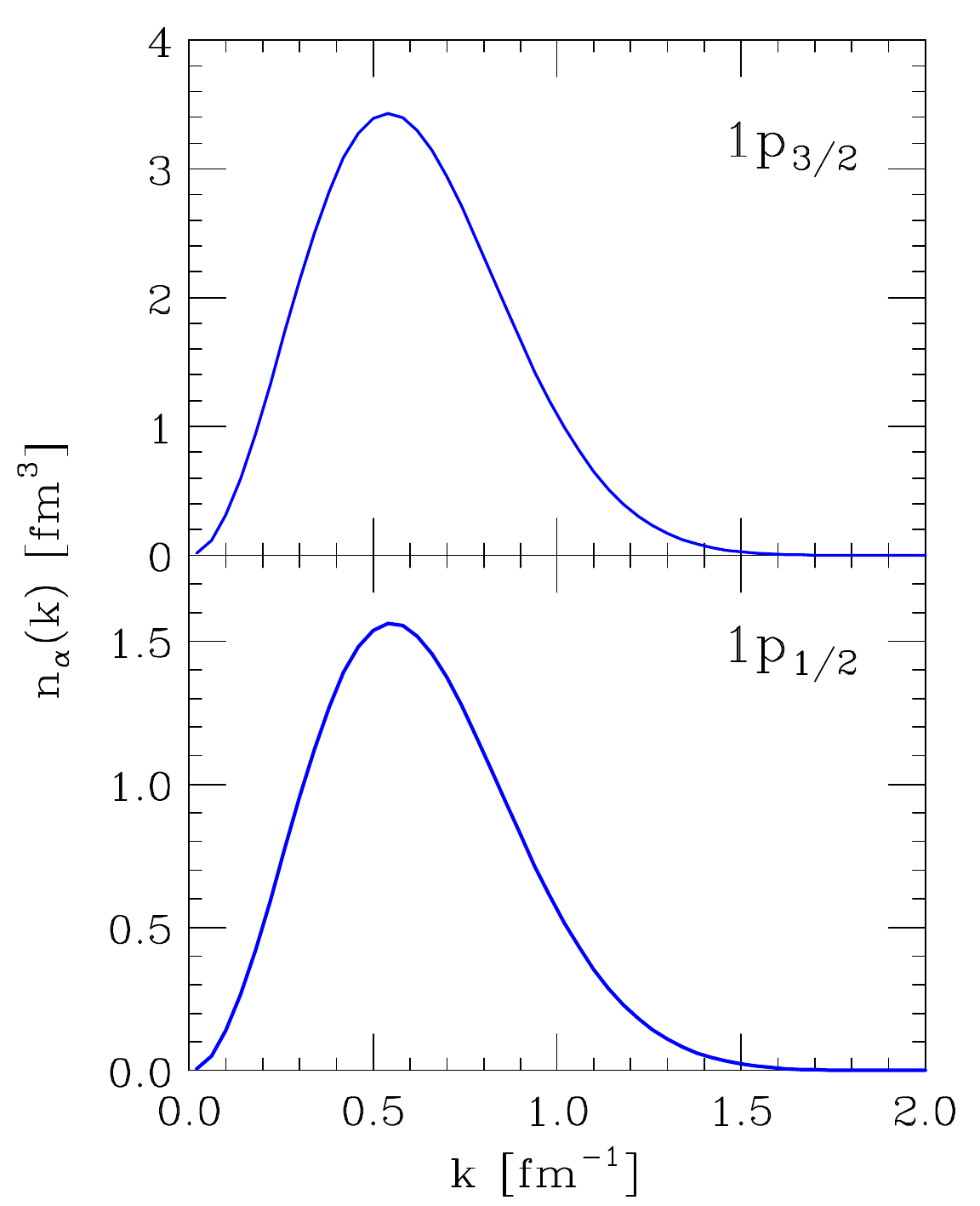}  \includegraphics[scale= 0.46]{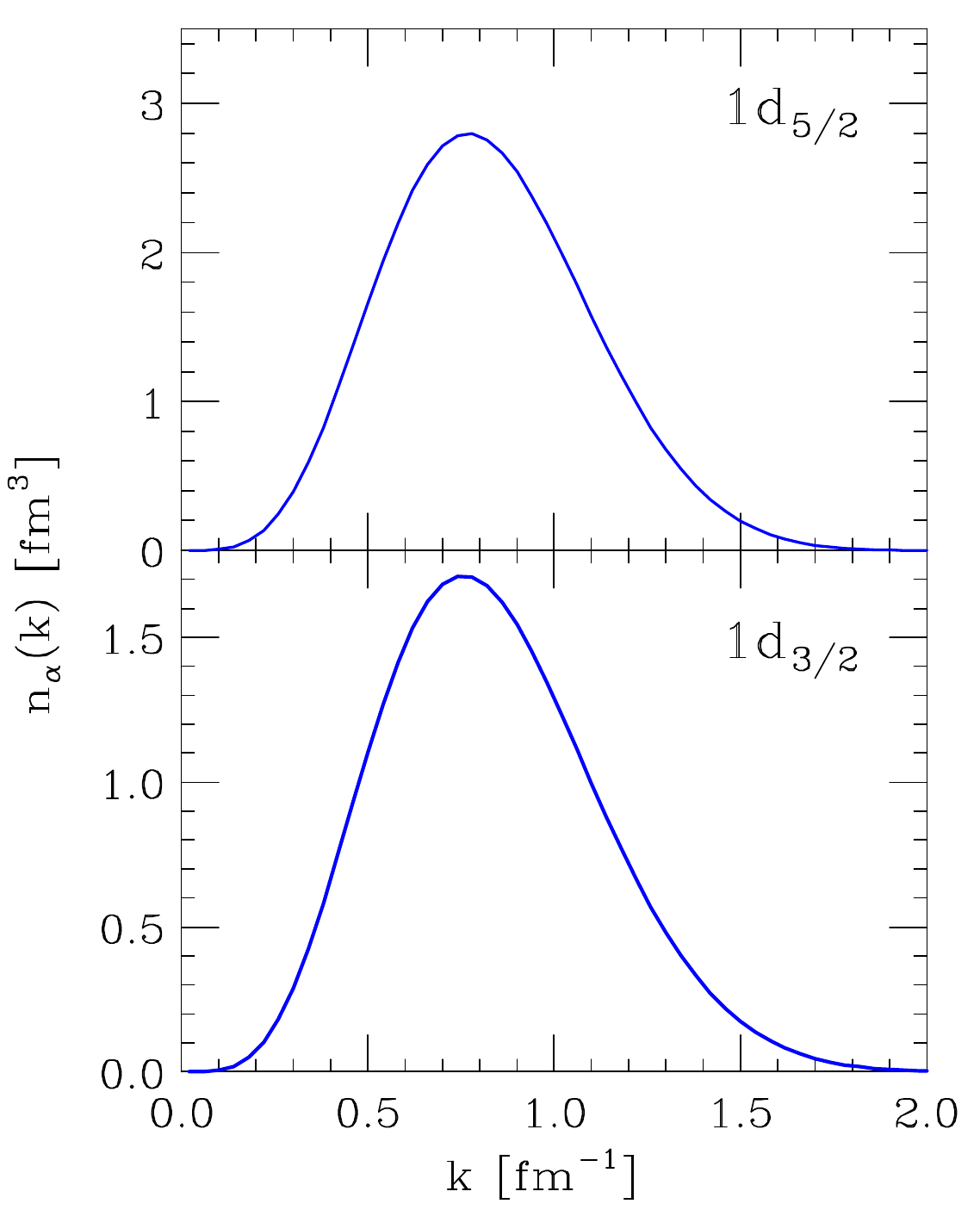}
\caption{Momentum distributions of the shell model orbits occupied by protons in the $\isotope[40][20]{Ca}$ ground state. The results have been
obtained using the Woods-Saxon potential reported in Ref.~\cite{bisconti}.}
\label{nk_shell}
\end{figure}

In the $\isotope[40][18]{Ar}$ ground state, protons occupy the $1s_{1/2}, \ 2s_{1/2}, \ 1p_{1/2}, \ 1p_{3/2}, \ 1d_{3/2}$ and $1d_{5/2}$ orbits. The corresponding momentum distributions are defined as
\begin{align}
\label{def:momdis}
n_\alpha(k) =  | \hat{\phi}_\alpha(k) |^2 \ ,
\end{align}
where $\alpha \equiv (n,\ell,j)$ denotes the set of quantum numbers specifying the shell model states and $\hat{\phi}_\alpha$ is
the Fourier transform of the corresponding radial wave function. The momentum distributions of the shell model orbits occupied by protons in the $\isotope[40][20]{Ca}$ ground state are illustrated in Fig.~\ref{nk_shell}. Note that the $n_\alpha(k)$ are normalized according to 
\begin{align}
\label{momdis:norm}
\int \frac{d^3 k}{(2 \pi)^3} \ n_\alpha(k) = N_\alpha \ ,
\end{align}
where $N_\alpha = 2j+1$ being the number of protons in the state $\alpha$.

The numerical results have been obtained through numerical solution of the Scr\"odinger equation with the Woods-Saxon potential reported in Ref.~\cite{bisconti} and the parameter set corresponding to $\isotope[40][20]{Ca}$.

{\bf Figure~\ref{nk_shell} shows that the missing momentum range relevant to the proposed study extends to $p_m \sim$ 350~MeV.}

%%%%%%%%%%%%%%%%%%%%%%%%%%%%%%%%%%%%%%%%%%%%%%%

%%%%%%%%%%%%%%%%%%%%%%%%%%%%%%%%%%%%%%%%%%%%%%%%%%%%%%%%%%%%%%%%%%%%%%%%%%%%%%%%%

\clearpage

\section{Kinematics and Experimental Rates}

In the past,  several $A(e,e^\prime p)$ experiments have been very successfully carried out in Hall A, using the HRS spectrometers. Measuring 
coincidence cross sections to obtain information on nuclear structure and dynamics  was in fact one of the primary goals of the early 
Physics program at CEBAF (see, e.g. Ref.\cite{mougey}). A list of  past Hall A experiments can be found in Tab.~\ref{tab:previous}. 
\begin{table}[!ht]
\caption{Past $A(e,e^\prime p)$ experiments in Hall A}
\begin{center}
\begin{tabular}{|l|l|}
\hline
 E89-003 ~& Study of the Quasielastic $(e,e^\prime p)$ reaction in $^{16}$O at High Recoil Momentum \cr  \hline
 E89-044 ~& Selected Studies of the $^3$He and 4He Nuclei through \ldots \cr \hline
 E97-111 ~& Systematic Probe of Short-Range Correlations via the Reaction $^4$He$(e,e^\prime p)^3H$ \cr \hline
 E00-102 ~& Testing the limits of the Single Particle Model in $^{16}$O($e,e^\prime p$) \cr \hline
 E03-104 ~& Probing the Limits of the Standard Model of Nuclear Physics with the $^4$He($e,e^\prime p$)$^3$H Reaction \cr \hline
 E04-004 ~& In-Plane Separations and High Momentum Structure in d($e,e^\prime p$)n \cr \hline
 E06-007 ~& Impulse Approximation limitations to the ($e,e^\prime p$) on $^{208}$Pb, $\ldots$\cr
 \hline
\end{tabular}
\end{center}
\label{tab:previous}
\end{table}

Throughout the experiments listed in Tab.~\ref{tab:previous} the knowledge on the technical parameters of the spectrometers has been continually improved, and the MC simulation packages that are now available have been refined and improved exploiting  the experience accumulated over the years.

The proposed  measurement of  the reaction Ar$(e,e^\prime p)$ will make use of the following key elements:
\begin{itemize}
\item the techniques that have been developed, and the lessons learned from the past experiments;
\item the Hall A high resolution spectrometers - HRSs;
\item the long experience with cryogenic high pressure targets in both Hall A and Hall C.
\end{itemize}

\textbf{The cross sections, the count rate calculations, and the simulations reported in this proposal were estimated using a beam current of 100~$\mu$A on a 10~Atm, 15~cm long Argon target running at a temperature of 130~K. The target density will be of of 560~mg~cm$^{-2}$. 
These setting parameters will be used with a \textbf{target luminosity of 5.5$\times$10$^{36}$ cm$^{-2}$~sec$^{-1}$}}.

The MCEEP software package was used to simulate the experiment,  with the following assumptions for each of the HRS spectrometers:
\begin{itemize}
\item $\dfrac{dp}{p}:$ 3.5\%
\item $\phi:~\pm~$20~mr
\item $\theta:~\pm~$40~mr.  
\end{itemize}
The above  assumptions are nominally used for cross section measurements with the HRS in Hall A.

\begin{figure}[htb!]
\label{example-data}
\centerline{\includegraphics[width=0.5\textwidth]{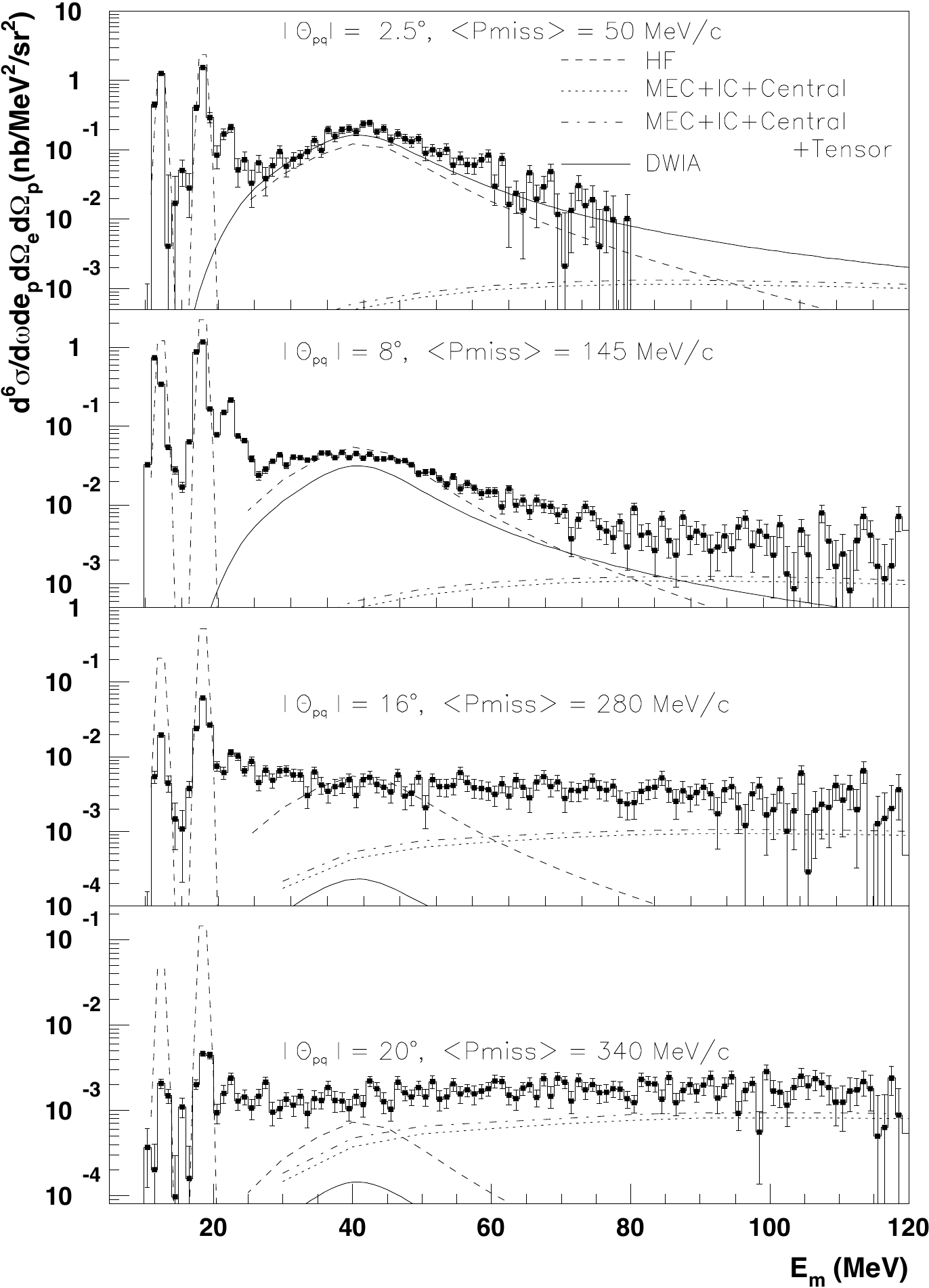}}
\caption{The E89-003 ran at a beam energy of 2.4~GeV and HRS settings very close to those proposed herein.  The EM resolution of 0.9~MeV FWHM was achieved and both bound state and continuum data was published in Physical Review Letters~\cite{Gao:2000ne,Liyanage:2000bf}.}
\end{figure}

We regard these settings and the experimental simulation as solid,  since they have already been proved to work excellently in the past.  The $^{16}$O($e,e^\prime p$) experiment E89-003, which ran at a beam energy of 2.4~GeV and HRS settings very similar to those proposed herein, reached an experimentally demonstrated $E_m$ resolution of 0.9~MeV FWHM (i.e. sigma of 0.38~MeV)~\cite{Gao:2000ne}. The data collected over the range of $E_m$ covered by the HRS are shown in Fig.~\ref{example-data}. 
The results of the E89-003 experiment have been published in Physical Review Letters as two separate articles~\cite{Gao:2000ne,Liyanage:2000bf}.

%%%%%%%%%%%%%%%%%%%%%%%%%%%%%%%%%%%%%%%%%%%%%%%%%%%%%%

\subsection{Experimental Setup}

\paragraph*{\textbf{Beam}}
\noindent We do not require beam polarization. The predicted count rates are such that the beam current can be less than 100~$\mu$A, although we have used a conservative estimate. The standard Hall A beam position and beam current monitors are more than adequate for our needs. The beam must be raftered over 2$\times$2~mm$^2$ on the cryogenic targets.

\vspace{0.5cm}

\paragraph*{\textbf{Target}} 
The parameters for the proposed 15 cm\footnote{We are looking into the possibility of using a slightly longer target (20~cm), in order to push the end cap further out of the acceptance of the HSRs.} cryogenic argon target are given in Tab.~\ref{tab:targ}. The relevant studies of computational fluid dynamics have been carried by Dr. S. Covrig~\cite{CovrigPC2014}.  The results indicate that with a beam current of 100~$\mu$A  the average density loss will be just 3.9\%. After consulting with the JLAB target group, we have also considered the possibility of reducing the pressure in the target cell in exchange for a thinner cell windows, the latter being a possible source of background.

\begin{table}[!h]
\begin{center}
\begin{tabular}{ c |c |c| c| c | c |c }
\hline
Length & Diameter& Pressure& Temperature&Density & Rad. Length & End Wall Caps\\
\hline\hline
15 cm& 2$\pi$ cm& 10 ATM& 130 K& 560 mg/cm$^2$ &2.9\%& 0.0254 cm\\
\hline
\end{tabular}
\end{center}
\caption{Parameters of the proposed Argon cryogenic target.}\label{tab:targ}
\end{table}

\subsection{Background and ancillary measurements}

We require additional beam time for the end cap background contribution, spectrometer pointing, carbon and hydrogen elastic measurements which require the following  targets:
\begin{itemize}
\item LH$_2$ cell of the same length as the Ar target - 15~cm (density 18.43 mg/cm$^2$);
\item C foils and multi-foils for y-target calibration 10~mils - (density 60 mg/cm$^2$);
\item Al dummy target (two foils Al with matched radiation length to filled Ar) (200 mg/cm$^2$) - current will be in this case limited to 25~$\mu$A.
\end{itemize}

\textbf{The beam time for the background measurements is comprised in the 9 days of requested beam time, and it has been included in Tab.~\ref{bt}}.

Carbon pointing (single foil carbon runs) and Al dummy runs will be taken at each settings. We will match the radiation length of the filled target with the Al target (i.e. the Al in the dummy is thicker then the cells). The data taking in this case will be very fast, as it was in the past~\cite{Alcorn:2004sb}. A preferred method will be fitting the shape of the Al data and subtracting the function with respect to just doing a data subtraction. In any case, we typically cut in y-target due to the very nice y-target resolution of the HRS ($\sim$1~mm \@ 90~ degrees) so this typically ends up being a small correction.

\subsection{Proposed Kinematics}

In the parallel kinematics of this experiment the electron beam energy, the final electron energy (and therefore the energy transfer) and the knocked out proton momentum are held fixed. In order to sample different missing momentum ($p_m$) the electron scattering angle is varied from approximately 25 degrees to 13 degrees, as shown in detail in Tab.~\ref{KinSelCoin1}. Note that, on account of the spectrometer acceptances, the kinematical setups of Tab.~\ref{KinSelCoin1}, corresponding to E$_m \approx$ 50~MeV, automatically cover all the relevant 
missing energy range.

\textbf{We are also requesting to take two settings in anti-parallel kinematics}, shown in Tab.~\ref{KinSelCoin2} as a cross-check of the spectral functions, or alternatively the momentum distributions, $n(k)$, extracted from the data. Being quantities describing intrinsic properties of the target, they should be independent of kinematics. 

Measuring the momentum distribution twice will also allow us to gauge the accuracy of the treatment of FSI within the approach of Ref.~\cite{FSI}.

%%%%%%%%%%%%%%%%%%%%%%%%%%%%%%%%%%%%%%%%%%%%%%%%%%%%%%%%%%%%%%%%%%%%%%%%%%%%%%%%%
\begin{table}[!ht]
\centering
\resizebox{\columnwidth}{!} {%
\begin{tabular}{ c | c c l c c c c c | c c}
\hline\hline
\multicolumn{11}{c}{
  Parallel Kinematics,
  Luminosity = $5.45\times 10^{36} {\rm atoms}~{\rm cm^{-2}}~{\rm sec^{-1}}$
} \\
\hline
kinematics & $E_e$ & $E_{e^\prime}$ & $\theta_e$  & $P_p$ & $\theta_p$ & $|{\bf q}|$
 & $p_m$ & $x_{bj}$ & $d\sigma_{eA}$ & Coin. Rate \\
% & & & & & & & & & & \\
 & ${\rm MeV}$ &  ${\rm MeV}$ & ${\rm deg}$ &${\rm MeV/c}$ & ${\rm deg}$
 & ${\rm MeV/c}$ & ${\rm MeV/c}$ &  & ${\rm mb}\,{\rm sr }^{-2}\,{\rm MeV}^{-2}$  & Hz \\
\hline\hline
   kin01 & 2200  &  1717  & 25.3  & 1000  &  -48.6  & 980          &   20         &  0.80  &  $0.173\times 10^{-6}$   & 0.60  \\
   kin02 & 2200  &  1717  & 23.9  & 1000  &  -47.8  & 940          &   60         &  0.72  &  $0.191\times 10^{-6}$   & 0.67  \\
   kin03 & 2200  &  1717  & 22.5  & 1000  &  -47.0  & 900          &   100       & 0.64   &  $0.214\times 10^{-6}$   & 0.75  \\
   kin04 & 2200  &  1717  & 21.1  & 1000  &  -45.9  & 860          &   140       & 0.56   &  $0.245\times 10^{-6}$   & 0.88  \\
   kin05 & 2200  &  1717  & 19.6  & 1000  &  -44.7  & 820          &   180       & 0.49   &  $0.238\times 10^{-6}$   & 0.85  \\
   kin06 & 2200  &  1717  & 18.1  & 1000  &  -43.2  & 780          &   220       & 0.41   &  $0.181\times 10^{-6}$   & 0.64  \\
   kin07 & 2200  &  1717  & 16.6  & 1000  &  -41.4  & 740          &   260       & 0.35   &  $0.107\times 10^{-6}$   & 0.39  \\
   kin08 & 2200  &  1717  & 15.0  & 1000  &  -39.3  & 700          &   300       & 0.28   &  $0.516\times 10^{-7}$   & 0.18  \\
   kin09 & 2200  &  1717  & 13.3  & 1000  &  -36.7  & 660          &   340       & 0.22   &  $0.250\times 10^{-7}$   & 0.09  \\
   kin10 & 2200  &  1717  & 11.5\footnote{For this one extreme kinematics, the electron arm will not be centered as it is limited to 12.5 degrees,
never the less, the indicated angle is within the acceptance.}  & 1000  &  -33.4  & 620          &   380       & 0.17   &   $0.171\times 10^{-7}$  & 0.06  \\
\hline \hline
\end{tabular}
}
\caption{\label{KinSelCoin1}Kinematical variables, PWIA $^{40}${Ca}$(e,e^\prime p)$
  cross section and coincidence rates in parallel kinematics. }
\end{table}
%
%%%%%%%%%%%%%%%%%%%%%%%%%%%%%%%%%%%%%%%%%%%%%%%%%%%%%%%%%%%%%%%%%%%%%%%%%%%%%%%%%
\begin{table}[!ht]
\centering
\resizebox{\columnwidth}{!}{%
\begin{tabular}{ c | c c c c c c c c | c c }
\hline\hline
\multicolumn{11}{c}{
  Anti-parallel Kinematics,
  Luminosity = $5.45\times 10^{36} {\rm atoms}~{\rm cm^{-2}}~{\rm sec^{-1}}$
} \\
\hline
Kinematics & $E_e$ & $E_{e^\prime}$ & $\theta_e$  & $P_p$ & $\theta_p$ & $|{\bf q}|$
 & $p_m$ & $x_{bj}$ & $d\sigma_{eA}$ & Coin. Rate \\
% & & & & & & & & & & \\
 & ${\rm MeV}$ &  ${\rm MeV}$ & ${\rm deg}$ &${\rm MeV/c}$ & ${\rm deg}$
 & ${\rm MeV/c}$ & ${\rm MeV/c}$ &  & ${\rm mb}\,{\rm sr }^{-2}\,{\rm MeV}^{-2}$  & Hz \\
\hline\hline
kin11 & 2200  & 1717   & 29.8  & 1000  &  -50.2      &  1110       &   -110     & 1.1  & $0.364\times 10^{-7}$   & 0.13  \\
kin12 & 2200  & 1717   & 34.4  & 1000  &  -51.1      &  1247       &   -247     & 1.5  & $0.211\times 10^{-8}$   & 0.01  \\
\hline \hline
\end{tabular}
}
\caption{\label{KinSelCoin2}Kinematical variables, PWIA $^{40}${Ca}$(e,e^\prime p)$
  cross section and coincidence rate in anti-parallel kinematics. }
\end{table}
%%%%%%%%%%%%%%%%%%%%%%%%%%%%%%%%%%%%%%%%%%%%%%%%%%%%%%%%%%%%%%%%%%%%%%%%%%%%

\subsection{Coincidence rates}

Tables~\ref{KinSelCoin1} and \ref{KinSelCoin2} provide a detailed account of the kinematic setups that we are proposing, along with the coincidence rates 
obtained from  the six-fold PWIA cross-sections  computed using the calcium spectral function of Ref.\cite{ANK} and the CC1 off shell extrapolation of the electron-nucleon cross 
section of Ref.~\cite{defo}.

\subsection{Singles rates}
The single-arm background rates, for both the parallel and antiparallel kinematics, are summarized  in Tab.~\ref{acc1} and~\ref{acc2}.

The $(e,e')$ rate were calculated  with the QFS computer code of Lightbody and O'Connell~\cite{Lightbody}, while the $(e, p)$, $(e, \pi^+)$, and 
$(e, \pi^-$) rates were obtained using  the EPC code, also from Ref.~\cite{Lightbody}. The rate calculations assumed a luminosity of $5.5\times 10^{36}
$ cm$^2$ s~$^{-1}$.

As demonstrated by many past experiment, our predicted rates have never been a problem for the Hall A apparatus, and can be easily achievable.

%%%%%%%%%%%%%%%%%%%%%%%%%%%%%%%%%%%%%%%%%%%%%%%%%%%%%%%%%%%%%%%%%%%%%%%%%%%%%%%%%
\begin{table}[!hb]
\begin{center}
%\resizebox{\columnwidth}{!}{%
\begin{tabular}{ c |  c c | c c }
\hline
\hline
\multicolumn{5}{c}{Luminosity = $5.45\times 10^{36} {\rm atoms}~{\rm cm^{-2}}~{\rm sec^{-1}}$} \\
\hline
 Kinematics & $d\sigma_{ee'}$ & Single Rate & $d\sigma_{e\pi^{-}}$ & Single Rate \\
 & & (e,e') & & (e,$\pi^{-}$) \\
 & ${\rm mb} \ {\rm sr }^{-1} \ {\rm MeV}^{-1}$ & Hz
 & ${\rm mb} \ {\rm sr }^{-1} \ {\rm MeV}^{-1}$ & Hz \\
\hline\hline
\multicolumn{5}{c}{Parallel Kinematics} \\\hline
   kin01 & $0.346\times 10^{-5}$ & 94.3 & $0.400\times 10^{-4}$ & 1090.0 \\
   kin02 & $0.516\times 10^{-5}$ & 140.6 & $0.467\times 10^{-4}$ & 1272.6 \\
   kin03 & $0.741\times 10^{-5}$ & 201.9 & $0.558\times 10^{-4}$ & 1520.6 \\
   kin04 & $0.106\times 10^{-4}$ & 288.9 & $0.684\times 10^{-4}$ & 1863.9 \\
   kin05 & $0.156\times 10^{-4}$ & 425.1 & $0.864\times 10^{-4}$ & 2354.4 \\
   kin06 & $0.246\times 10^{-4}$ & 670.4 & $0.113\times 10^{-3}$ & 3079.3 \\
   kin07 & $0.404\times 10^{-4}$ & 1100.9 & $0.154\times 10^{-3}$ & 4196.5 \\
   kin08 & $0.677\times 10^{-4}$ & 1844.9 & $0.221\times 10^{-3}$ & 6022.3 \\
   kin09 & $0.115\times 10^{-3}$ & 3133.8 & $0.337\times 10^{-3}$ & 9183.3 \\
   kin10 & $0.201\times 10^{-3}$ & 5477.3 & $0.562\times 10^{-3}$ & 15324.5 \\
\hline
\multicolumn{5}{c}{Anti-parallel Kinematics} \\\hline
   kin11 & $0.670\times 10^{-6}$ & 18.4 & $0.272\times 10^{-4}$ & 741.2 \\
   kin12 & $0.986\times 10^{-7}$ & 2.7 & $0.350\times 10^{-4}$ & 953.8 \\
\hline \hline
\end{tabular}
%}
\caption{\label{acc1} $(e,e^\prime)$ and $(e,\pi^{-})$ single electron-arm rates along with accidental rates.}
\end{center}
\end{table}
%%%%%%%%%%%%%%%%%%%%%%%%%%%%%%%%%%%%%%%%%%%%%%%%%%%%%%%%%%%%%%%%%%%%%%%%%%%%%%%%%
\begin{table}[!hb]
\begin{center}
%\resizebox{\columnwidth}{!}{%
\begin{tabular}{ c | c c | c c }
\hline
\hline
\multicolumn{5}{c}{Luminosity = $5.45\times 10^{36} {\rm atoms}~{\rm cm^{-2}}~{\rm sec^{-1}}$} \\
\hline
 Kinematics & $d\sigma_{ep}$ & Single Rate & $d\sigma_{e\pi^{+}}$ & Single Rate \\
 & & (e,p) & & (e,$\pi^{+}$) \\
 & ${\rm mb} \ {\rm sr }^{-1} \ {\rm MeV}^{-1}$ & Hz
 & ${\rm mb} \ {\rm sr }^{-1} \ {\rm MeV}^{-1}$ & Hz \\
\hline\hline
\multicolumn{5}{c}{Parallel Kinematics} \\\hline
   kin01 & $0.137\times 10^{-3}$ & 3733.2 & $0.378\times 10^{-4}$ & 1030.0 \\
   kin02 & $0.158\times 10^{-3}$ & 4305.5 & $0.443\times 10^{-4}$ & 1207.2 \\
   kin03 & $0.183\times 10^{-3}$ & 4986.8 & $0.531\times 10^{-4}$ & 1447.0 \\
   kin04 & $0.217\times 10^{-3}$ & 5913.3 & $0.654\times 10^{-4}$ & 1782.2 \\
   kin05 & $0.260\times 10^{-3}$ & 7085.0 & $0.828\times 10^{-4}$ & 2256.3 \\
   kin06 & $0.318\times 10^{-3}$ & 8665.5 & $0.109\times 10^{-3}$ & 2970.3 \\
   kin07 & $0.397\times 10^{-3}$ & 10818.3 & $0.148\times 10^{-3}$ & 4033.0 \\
   kin08 & $0.509\times 10^{-3}$ & 13870.3 & $0.213\times 10^{-3}$ & 5804.3 \\
   kin09 & $0.675\times 10^{-3}$ & 18393.8 & $0.327\times 10^{-3}$ & 8910.8 \\
   kin10 & $0.931\times 10^{-3}$ & 25369.8 & $0.546\times 10^{-3}$ & 14878.5 \\
\hline\hline
\multicolumn{5}{c}{Anti-parallel Kinematics} \\\hline
   kin11 & $0.948\times 10^{-4}$ & 2583.3 & $0.255\times 10^{-4}$ & 694.9 \\
   kin12 & $0.121\times 10^{-3}$ & 3297.3 & $0.330\times 10^{-4}$ & 899.2 \\
\hline \hline
\end{tabular}
%}
\caption{\label{acc2} $(e,\pi^{+})$ and $(e,p)$ single proton-arm rates.}
\end{center}
\end{table}
%%%%%%%%%%%%%%%%%%%%%%%%%%%%%%%%%%%%%%%%%%%%%%%%%%%%%%%%%%%%%%%%%%%%%%%%%%%%%%%%%
\subsection{Systematic and Statistical Uncertainties}
We have estimated a statistical uncertainty for our measurements of 3\%, and we assumed a systematic uncertainties of 3\%.

In order to achieve that we will require luminosity monitoring and elastic cross section measurements at appropriate scattering angles to test the counting rates as a function of the beam current. This was done in the past by other experiments and will guarantee systematic uncertainties of the order of 3\%, that will be added in quadrature to the statistical uncertainties.

\clearpage

\section{Summary and Beam Time Request}
\subsection{Summary}

The spectral function, trivially related to the two-point Green's function, is a fundamental quantity describing the dynamics of interacting many-body systems. As such, its measurement through the analysis of the nuclear $(e,e^\prime p)$ cross section is certainly valuable in its own right.

We propose to carry out an experiment aimed at obtaining the argon spectral function. The results of this measurement, besides yielding previously unavailable information on nuclear structure and dynamics, will provide the input needed to improve the simulation of neutrino interactions in liquid argon detectors, thus reducing the systematic uncertainty associated with the oscillation analysis of experiments such as LBNE.

The proposed measurement is mainly focused on the energy and momentum region in which shell model dynamics is known to dominate. The understanding of this region will allow for accurate prediction of the CC QE cross section in the single-nucleon knock out sector, which is known to be of paramount importance. However, it is important to keep in mind the that spectral function provides a description of the target ground state. Therefore, its knowledge is needed for the description of all interactions involving a single nucleon, independent of the final state.

Theoretical studies and extensive comparison with electron scattering data~\cite{LDA,PRD,ANK2} suggest that highly realistic models of the full spectral functions, suitable to describe two-nucleon emission processes arising from short range  correlations in the initial state can be obtained combining the measured mean field spectral function with theoretical calculations of the continuum component within the LDA scheme~\cite{LDA}. 

The impact of the nuclear spectral function on neutrino oscillation parameters has been recently studied in Ref.~\cite{gofsix}. 

The results of this exploratory analysis, carried out using the GLoBES sensitivity framework,  clearly show that the description of the nuclear ground state has a {\em non-negligible} influence on the determination of the atmospheric oscillation parameters in a typical $\nu_\mu$ disappearance experiment. 

\subsection{Beam Time Request}

\textbf{To perform this experiment and make the necessary systematic measurements  we request 9 days of beam time at 2.2~GeV as shown in Tab.~\ref{bt}}. 

Time is also included to take carbon pointing data and multi-foil carbon target data to allow a calibration of the pointing of the spectrometers and y-target calibrations. The carbon data will also be useful for obtaining an independent measurement of  the absolute missing mass calibration. The Al dummy data will allow us to subtract the end caps of the target cell from the data. 

%
%%%%%%%%%%%%%%%%%%%%%%%%%%%%%%%%%%%%%%%%%%%%%%%%%%%%%%%%%%%%%%%%%%%%%%%%%%%%%%%%%
\begin{table}[!htb]
\begin{center}
%\resizebox{\columnwidth}{!}{%
\begin{tabular}{ c | c}
\hline
\hline
\multicolumn{2}{c}{
Luminosity = $5.45\times 10^{36} {\rm atoms} ~{\rm cm^{-2}} ~{\rm sec^{-1}}$} \\\hline
Kinematics & Beam Time \\
  & day(s) \\
\hline
\multicolumn{2}{c}{Parallel Kinematics} \\\hline
   kin1    & 0.25 \\
   kin2    & 0.25 \\
   kin3    & 0.25 \\
   kin4    & 0.25 \\
   kin5    & 0.25 \\
   kin6    & 0.25 \\
   kin7    & 0.25 \\
   kin8    & 0.25 \\
   kin9    & 0.50 \\
   kin10  & 1.00 \\\hline
\multicolumn{2}{c}{Anti-parallel Kinematics} \\\hline
   kin11  & 0.50 \\
   kin12  & 3.00 \\
\hline \hline
\multicolumn{2}{c}{Calibration (Beam energy, current, etc.)} \\\hline
   cal. & 1.0 \\\hline\hline
\end{tabular}
%}
\caption{\label{bt} Required beam time in  parallel and antiparallel kinematics, and calibration. }
\end{center}
\end{table}
%%%%%%%%%%%%%%%%%%%%%%%%%%%%%%%%%%%%%%%%%%%%%%%%%%%%%%%%%%%%%%%%%%%%%%%%%%%%%%%%%
\newpage
\clearpage

%%%%%%%%%%%%%%%%%%%%%%%%%%%%%%%%%%%%%%%%%%%%%%%%%%%%%%%%%%%%%%%%%%%%%%%%%%%%%

\end{document}